\documentclass[twocolumn]{aastex631}

\usepackage{amsmath}

\begin{document}

\title{Oort Cloud Formation and Evolution in Star Clusters}

\author{Justine C. Obidowski}
\affiliation{David A. Dunlap Department of Astronomy \& Astrophysics, University of Toronto, 50 St. George St, Toronto, ON M5S 3H4, Canada}
\affiliation{Department of Physics and Astronomy, University of British Columbia, Vancouver, BC V6T 1Z1, Canada}
\email{jusdowski@phas.ubc.ca}

\author{Jeremy J. Webb}
\affiliation{David A. Dunlap Department of Astronomy \& Astrophysics, University of Toronto, 50 St. George St, Toronto, ON M5S 3H4, Canada}
\affiliation{Department of Science, Technology and Society, Division of Natural Science, York University, Toronto, ON M3J 1P3, Canada}
\email{webbjj@yorku.ca}

\author{Simon Portegies Zwart}
\affiliation{Leiden Observatory, Leiden University, P.O. Box 9513, 2300 RA, Leiden, The Netherlands}

\author{Maxwell X. Cai}
\affiliation{Leiden Observatory, Leiden University, P.O. Box 9513, 2300 RA, Leiden, The Netherlands}

\begin{abstract}
It is unknown if an Oort cloud reaches its maximum mass within its star’s birth cluster or millions of years later. Complicating the Oort cloud evolution process is the fact that comets can be stripped from orbit due to perturbations from passing stars. We explore how a star’s cluster escape time (t$_{ \rm esc}$) and the time its Oort cloud reaches maximum mass (t$_{ \rm max}$) affect the Oort cloud's ability to survive via $N$-body simulations. In a 14 M$_\odot$/pc$^3$ cluster, we identify 50 stars of 1 M$_\odot$ with a range of t$_{ \rm esc}$ to host Oort clouds, each with 1000 comets at t$_{ \rm max}$. For each host, we consider Oort clouds that reach maximum mass 0, 50, and 250 Myr after the cluster's formation. Each Oort cloud’s evolution is simulated in the cluster from t$_{ \rm max}$ to t$_{ \rm esc}$. Only a fraction of comets tend to remain in orbit, with this amount depending on t$_{ \rm max}$ and t$_{ \rm esc}$. We observe that 12\%, 22\%, and 32\% of Oort clouds with a t$_{ \rm max}$ of 0, 50 and 250 Myr retain $>$50\% of their comets at t$_{ \rm esc}$, respectively. We find that the fraction of comets stripped has the relationship, $\rm f=m\log_{10}(\frac{t_{ \rm esc}-t_{ \rm max}}{Myr})$ where m = 0.32$\pm$0.04, indicating that the longer the Oort cloud remains in the cluster, the more comets are stripped, with this fraction increasing logarithmically at approximately the same rate for each t$_{ \rm max}$.

\end{abstract}


\section{Introduction} \label{sec:intro}
In the outermost regions of a planetary system, a spherical cloud of millions of cometary objects is likely to exist in orbit around the star. This distant cloud is known as the Oort cloud, named after Dutch astronomer Jan Hendrik Oort, who first hypothesized its existence in the Solar System in 1950 \citep{Oort1950}. Although no confirmed direct observations of the Solar System’s Oort cloud or any exo-Oort cloud have been made, there are strong indications that one exists around the Sun. One important piece of evidence of the Oort cloud's existence in the Solar System is long-period comets, which are predicted to have been perturbed from the Oort cloud and sent on highly elliptical orbital paths that pass close to the Sun. Long-period comets have periods of hundreds or thousands of years and have been observed throughout history, but more of them have been observed and classified in recent decades due to improvements in observing technology \citep{Kr_likowska_2019}. Another sign of the existence of the Oort cloud is the discovery of distant transneptunian objects (TNOs) that have aphelions near the inner Oort cloud. For example, several TNOs have been observed to have highly eccentric orbits with a perihelion beyond the Kuiper cliff at 47.8 AU \citep{Bannister2017, Sheppard2019, Alexandersen2019}, such as Sedna \citep{Brown2004} and 2012 VP113 \citep{Trujillo2014}. 

The Oort cloud of the Sun is split into an inner and outer region. The inner Oort cloud is considered to be the region beyond the Kuiper belt that extends from approximately 2,000 to 10,000 AU while the outer Oort cloud ranges from approximately 10,000 to 200,000 AU. The outer boundary of an Oort cloud defines the cosmographic boundary of the planetary system and the extent of the Hill sphere of the host star \citep{Hill1913}. For the Sun, the outer boundary of the Oort cloud is the distance equal to the Sun's Hill sphere radius in the Galactic potential ($\sim$200,000 AU) \citep{Chebotarev1965}. The inner boundary of an Oort cloud is defined less clearly and is still debated for the Solar System \citep{Hills1981, Leto2008, Brasser2015}. 

Stars do not form in isolation, but in clustered environments of gas and dust within a galaxy \citep{Lada2003, Simon2009, Adams2010}. Within these clustered environments, star formation can occur and result in the birth of a large group of stars known as a star cluster. The evolution of bodies in orbit around a star within a star cluster such as cometary objects in an Oort cloud can be affected by perturbations from other stars passing close to the system. A field star can also experience perturbations caused by another star, but they are different than those experienced in a cluster. Perturbations from passing stars depend on both the stellar density of their environment and the effective cross-section for the perturbation to occur \citep{Jilkova2016}. This cross-section decreases rapidly with increasing interaction speed, which is different in the cluster (a few km/s) than in the field ($\sim$40 km/s). If a star is located in a dense cluster of stars, it will likely experience a high number of close encounters with other stars that can cause perturbations. Objects in orbit close to their host star are usually less affected by perturbations from passing stars as they are more strongly held by their star's gravity compared to objects that are further away. Studies of planetary systems in star clusters \citep{Brasser2012, Brasser2015, Cai2017b, Brown2022} have shown that perturbations from passing stars can change the orbits of objects in a system and can even cause some objects, such as comets in an Oort cloud, to be stripped from orbit entirely. Objects stripped from their orbits around the Sun that do not get captured by another body become free-floating interstellar objects as they are ejected from the Solar System. 

 There are two popular scenarios for the formation of the Oort cloud. The first one explains that the Oort cloud is formed mainly by ejecting inner Solar System material through planet–disk interactions \citep{Dones2000, Dones2004}, and the other involves formation through two separate processes: local disk asteroids getting ejected into an inner region at around 3,000-20,000 AU and free-floating debris from the birth star cluster accreting in an outer region ($>$20,000 AU) \citep{Zheng1990, Valtonen1992, Brasser2006, Brasser2007, Brasser2008}. The timescale of an Oort cloud’s formation is strongly connected to its star’s birth environment \citep{Brasser2012} and there are several possible timescales depending on the chosen formation model. Estimates of the timescale of the Oort cloud's formation range from almost instantaneously after the formation of the Sun \citep{Fernandez1997}, synchronously with Jupiter’s formation \citep{Stevenson1988, Fernandez2000, Dones2004}, after Jupiter's formation and migration \citep{Shannon2019}, slowly growing over several hundred Myr \citep{Kaib2008, Nordlander2017}, and growing over Gyr timescales \citep{Duncan1987}.

\cite{Simon2021} performed a numerical investigation that showed that during the Sun’s infancy, the Oort cloud could have formed from complex interactions between the Sun and nearby stars, free-floating debris, the Galactic tidal field, and planetary scattering. Specifically, they found that the Solar System’s birth environment prevents the gas giant planets from forming the Oort cloud, but still stimulates its formation by scattering outer regions of the protoplanetary disk and capturing new comets from other stars or the interstellar free-floating population (see \citealp{Zheng1990}). Simulations by \cite{Levison2010} also found that a large number of comets in the Oort cloud could be captured from the protoplanetary disks of other stars in the Sun’s birth cluster. According to this view, the Oort cloud has only a minor contribution from the asteroids scattered by the gas giant planets, Jupiter and Saturn. The ice giants, Uranus and Neptune were found to be more favorable in producing the Oort cloud (see \citealp{Leto2008, Paulech2010}) as they follow the same process of ejecting asteroids, but their wider orbits and lower mass result in a longer timescale with smaller changes in the semi-major axis and orbital eccentricity of the asteroids (see \citealp{Duncan1987, Fernandez2004, Correa2019}). The longer timescale of the ice giants ejecting asteroids is important for the formation of the Oort cloud because while the Sun is still a member of its birth cluster, comets in wide orbits are easily stripped from orbit due to perturbations \citep{Nordlander2017}. \cite{Simon2021} found that $\sim$70\% of the material in the Oort cloud originated from the region in the protoplanetary disk that was located between 15-35 AU (near the current location of the Centaur asteroids and ice giants). 

Over time, star clusters dissolve as stars escape the cluster. The time that a star leaves the cluster is its escape time (t$_{\rm esc}$), which is defined as when the star’s total energy with respect to the cluster's gravitational energy is greater than zero. For a host star with a given escape time, the amount of time that its Oort cloud will be subjected to external perturbations depends on how long it took the Oort cloud to build up to its maximum mass. If the Oort cloud builds up quickly such that it reaches maximum mass at approximately the same time as its host star forms, it would have to survive longer in the cluster than if it forms closer to the star's escape time. In addition, the young cluster is considerably denser compared to when its closer to dissolution. 

Can an Oort cloud that forms around a star within a star cluster leave the cluster with most of its comets remaining in orbit?  To explore this question and the implications of the Oort cloud formation theories, our research explores the co-evolution of Oort clouds and a star cluster. We observe how a star’s escape time and the time it takes for an Oort cloud to reach its maximum mass affect the amount of comet stripping caused by perturbations from passing stars. To accomplish this task, $N$-body simulations of Oort clouds around various solar mass stars in a star cluster are performed and analyzed.

In this study, we explore general Oort cloud evolution in a star cluster that is not specific to only the Solar System. As the formation and evolution of Oort clouds is still being investigated, we will not explore the specific process of Oort cloud formation but we will consider different times an Oort cloud could reach maximum mass in its star's birth cluster. Exploring when an Oort cloud reaches this peak mass can inform us about its evolution in clusters at different density states. We will consider Oort clouds that reach maximum mass at the beginning of the star cluster's dense formation state, as well as Oort clouds that require 50 and 250 Myr to build up to this maximum mass, which is when the cluster is less dense due to stellar evolution and relaxation induced expansion.

In Section \ref{sec:maths} we outline our method for simulating Oort cloud comets around stars orbiting within a star cluster. We present the results of these simulations in Section \ref{sec:results}, where we explore what fraction of a star's Oort cloud remains bound after the star escapes its birth cluster. In Section \ref{sec:discussion}, we explain what our results mean for Oort clouds around Sun-like stars before summarizing our findings in Section \ref{sec:conclusions}.

\section{Methodology} \label{sec:maths}
\subsection{Simulation Methods}
To explore the evolution of an Oort cloud around a star in a clustered environment, we perform $N$-body simulations of hypothetical Oort clouds around stars within a star cluster. To perform the simulations, the Lonelyplanets method outlined in \cite{Cai2017} is used. The star cluster is simulated using the $N$-body simulation code, \texttt{NBODY6++GPU} \citep{Wang2015} which integrates the star cluster dynamics to produce the cluster’s dynamical evolution as a function of time. The closest stars (perturbers) to the chosen host star are identified throughout the cluster simulation. The output of the $N$-body simulation of the star cluster is then used to create an $N$-body simulation of an Oort cloud around the host star in the cluster. Each Oort cloud is simulated around its star and integrated over time using the $N$-body integrator \texttt{REBOUND} \citep{Rein2012}, saving time steps at least every 1000 years. The \texttt{Block Time Step} scheme \citep{Cai2015} is used to simulate the star cluster and Oort cloud separately which allows the dynamical timescale of the cluster to be a few orders of magnitude larger than that of the Oort cloud’s comets. In summary, the simulation code calculates the position and velocity of the host star, comets, and perturbers at time steps throughout the cluster’s evolution until the host star escapes the cluster. 

\subsection{Simulation Parameters}
\label{sec:maths2}
The star cluster used for the simulations is a typical open cluster that is comparable to estimates of the Sun's birth cluster \citep{Simon2009, Adams2010, Simon2019}. The cluster has an initial total mass of 7,500 M$_\odot$ consisting of 12,500 stars, an initial half-mass radius of 4 parsecs, and a circular orbit at 5 kilo-parsecs from the center of a Milky Way-type galaxy. Based on these properties, the initial density of the star cluster within the half-mass radius is 14 M$_\odot$/pc$^{3}$ . The cluster dissolves after $\sim$1 Gyr, which is determined as the time when less than 100 stars remain in the cluster. Although most stars form in clusters, only $\sim$10-30\% of stars are thought to form within clusters that evolve into open clusters (see \citealp{Vandenbergh1981, Elmegreen1985, Battinelli1991, Adams2001b, Kruijssen2012}). The remaining percentage of stars are born in clusters with much shorter lifetimes. The timescales for the disruption of these short-lived clusters are estimated to be only a few tens of millions of years, which is significantly shorter than estimates for the Sun's birth cluster. Thus, our work focuses only on solar mass stars that form within this long-lifetime cluster, as they could represent an evolution scenario of the Sun's Oort cloud. To best compare the results of our work to the Oort cloud of the Solar System, we select 50 of the 150 stars of 1 M$_\odot$ in the cluster as host stars for the Oort clouds. The 50 stars that are selected have a range of escape times as we want to study the effect of escape time on Oort cloud evolution.  The formation process of the stars in the clusters themselves is not considered as they are assumed to have already formed before the start of the simulation.  The simulations also do not consider the formation and evolution of any planets around the stars.

Each Oort cloud is simulated starting from its maximum mass, as we assume that its build-up has already occurred and that no additional mass will be added for the rest of the time it spends in the cluster.  The maximum mass of each Oort cloud is assumed to be 1000 comets, which are taken to be mass-less in the simulations such that they will only be affected by perturbations from passing stars and not each other. Once a comet is stripped from its Oort cloud, it is removed from the simulation. The comets have initial positions that are uniformly distributed around their host star with orbital eccentricities ranging from 0 to 0.99, with higher eccentricities being more common. The semi-major axes of comets in an Oort cloud initially range from 0.1 to 1 times their host star’s initial Hill sphere radius (R$_{\rm HS}$). As a result of the semi-major axis ranges, the majority of comets are given a perihelion greater than 1000 AU, although there are a few thousand comets across all simulations that have a perihelion less than this value with the minimum perihelion being 95 AU. 

The Hill sphere defines the effective gravitational sphere of influence around a star. The Hill sphere radius of a host star is determined based on its distance to the closest star and is calculated as
\begin{equation}
    \rm R_{HS}= d \left(\frac{m}{3M}\right)^{\frac{1}{3}}
	\label{eq:quadratic}
\end{equation}
where M is the host star's mass, d is the distance between the host star and the closest star and m is the closest star's mass. The gravitational potential of the cluster also provides a tidal force on the host star, defining a second type of Hill sphere that depends on the structure of the cluster and the star's location within it. We find that the Hill radius imposed by the cluster potential is on average within $\sim$20\% of the Hill radius imposed by the nearest neighboring star. Thus, we will only use equation 1 to define the Hill radius of each host star.

In our suite of simulations, we allow the time at which the Oort cloud reaches its peak mass to vary by initializing the Oort cloud simulation at different times. The maximum mass time (t$_{\rm max}$)  is the time it takes the Oort cloud to build up to its maximum mass relative to the initial/densest state of the star cluster. It is also the time that the Oort cloud simulation begins, allowing us to see what differs in the same Oort cloud's evolution when it reaches peak mass later in the star cluster's evolution. Preliminary simulations demonstrate that hardly any comets are stripped once a star escapes the cluster, meaning that we can end all simulations at the star's escape time. We simulate an Oort cloud around each host star starting from three different maximum mass times: 0, 50, and 250 Myr after the initial/densest state of the star cluster, and run until the star escapes the cluster. These times were selected to explore Oort cloud evolution in three different stages of the cluster’s life, observing how the cluster's decrease in density over time affects the Oort cloud. The maximum mass time of 0 Myr was chosen to represent the case of very rapid Oort cloud formation after the star’s birth. In the cases of the later times, a maximum mass time of 50 Myr represents a scenario where the Oort cloud built up to maximum mass recently after the planetary system’s formation was completed, while a time of 250 Myr represents a scenario of the Oort cloud taking millions of years after the planetary system’s formation to build up to maximum mass. Another possible interpretation of these three times is that they can each indicate t = 0 in ``different'' clusters, which will be less dense and less massive than our actual t = 0 cluster, allowing us to observe Oort cloud evolution in effectively different clusters.

\begin{figure}
  \includegraphics[width=\linewidth]{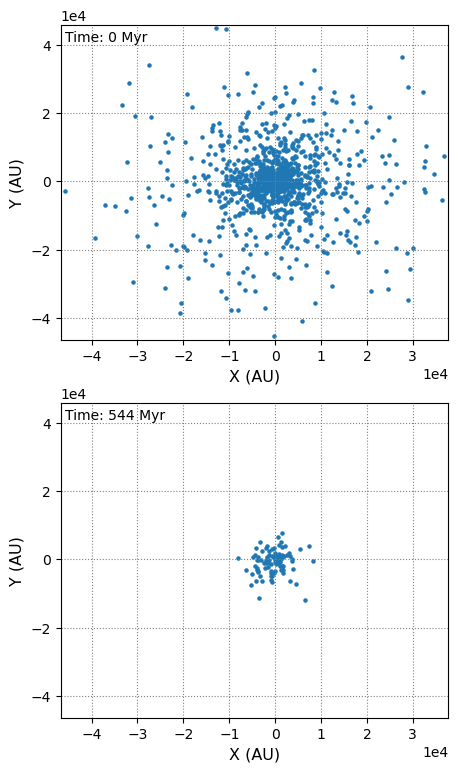}
  \caption{The initial (top) and final (bottom) positions of comets in the simulation of an Oort cloud that reached its maximum mass at the star cluster's initial state (t$_{\rm max}$ = 0). The blue dots represent comets in orbit around the star. The simulation begins with 1000 comets and ends with 93 comets remaining in orbit when the star leaves the cluster.}
  \label{fig:fig1}
\end{figure}

\section{Results}
\label{sec:results}
We present a total of 50 different sets of three simulations as the 50 host stars are simulated with an Oort cloud starting at the three maximum mass times. In each simulation, the number of comets in orbit and their characteristics are calculated at each time step, allowing this data to be analyzed over time until the host star leaves the cluster. We present the results of a single Oort cloud system in Section \ref{sec:results1}, where we explore the number of comets stripped from orbit over time due to perturbations from passing stars and how these perturbations have affected the orbits of the remaining comets. Furthermore in Section \ref{sec:results2}, we analyze the total fraction of comets stripped from orbit in each Oort cloud when the host star escapes the cluster.

\subsection{Walkthrough of a Single Star's Oort Cloud Evolution}
\label{sec:results1}

Figure \ref{fig:fig1} shows the initial and final positions of comets in the simulation of a single Oort cloud that reached its maximum mass at the star cluster's initial state (t$_{\rm max}$= 0 Myr). This particular host star has a cluster escape time of 544 Myr and this simulation is just one example of an Oort cloud that lost most (90.7\%) of its comets due to perturbations from passing stars.

\begin{figure}
  \includegraphics[width=\linewidth]{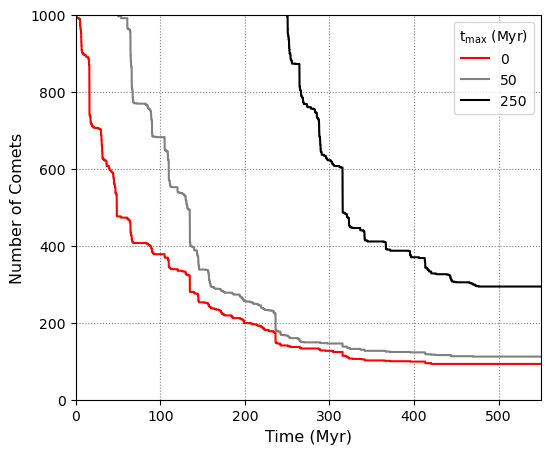}
  \caption{The number of comets in orbit for the Oort cloud simulations of one host star over time. This host star has an escape time of 544 Myr and loses most of its comets regardless of when its Oort cloud reaches maximum mass. The number of comets in orbit decreases over time for each simulation. Each drop in comet numbers is due to perturbations from passing stars which strip these comets from orbit, with sharp drops occurring when there has been a strong perturbation. }
  \label{fig:fig2}
\end{figure}

For the three simulations of the star's Oort cloud, Figure \ref{fig:fig2} shows the number of bound comets over time. This star loses the majority of its comets regardless of when its Oort cloud reaches maximum mass. The number of bound comets decreases over time for each of the three simulations. Each drop in comet numbers is due to perturbations from close encounters with passing stars in the cluster which strip these comets from orbit. In Figure \ref{fig:fig2}, we can see that the fraction of comets lost is proportional to the time an Oort cloud spends in the cluster after it has reached its maximum mass and inversely proportional to the time it takes the Oort cloud to build up this maximum mass. The 50\% dissolution time of an Oort cloud is the time it takes an Oort cloud to lose half of its maximum mass. For our single Oort cloud system shown in Figure \ref{fig:fig2}, its 50\% dissolution time is 50 Myr when t$_{\rm max}$ = 0, 130 Myr when t$_{\rm max}$ = 50 and 315 Myr when t$_{\rm max}$ = 250 Myr. These times indicate that the Oort cloud of this star loses half of its comets approximately 50-80 Myr after it reaches its maximum mass. 

The specific times when comets are stripped in Figure \ref{fig:fig2} show that more comets are stripped closer to when the Oort cloud is at its maximum mass, which is expected due to the cluster's higher density at this time, although density is not the only factor to consider. The level of perturbation an Oort cloud experiences not only depends on the density profile of the star cluster, but also on the path that its star travels in the cluster as it evolves. A perturbation from even just one passing star can affect the orbits of multiple bodies simultaneously. There are strong perturbations shown in Figure \ref{fig:fig2} that occur at early times that strip multiple comets at once, sometimes stripping hundreds at the same time. Many perturbations that occur at later times are weak and strip only a few comets. The three simulations also have a similar pattern of perturbations over time. This pattern is due to the evolution of the cluster, which is independent of the Oort cloud. The location of stars in the cluster over time is the same for each simulation and does not depend on when the Oort cloud reaches maximum mass. Therefore, every time an Oort cloud is simulated around the same star in the cluster, it would have the same stars pass by it at specific times, but this does not always affect the Oort cloud the same way. Our simulated comets are test particles, thus if one simulated test particle gets perturbed by the passing star, it indicates that any comets in that region of the Oort cloud (no matter if it's high or low in mass) will get perturbed. The evolutionary history of an Oort cloud is thus different if it experiences perturbations between 0-50 Myr or 0-250 Myr compared to if it does not. An Oort cloud that reaches its maximum mass earlier could be more susceptible to a later perturbation if it has already undergone some interactions that have caused it to expand. An Oort cloud that forms later wouldn't have experienced those early perturbations, so it may be more compact when the perturbation occurs. In contrast, a perturbed Oort cloud may have fewer weakly bound objects in the outermost fragile orbits, which makes its remaining comets harder to strip. 

\begin{figure}
  \includegraphics[width=\linewidth]{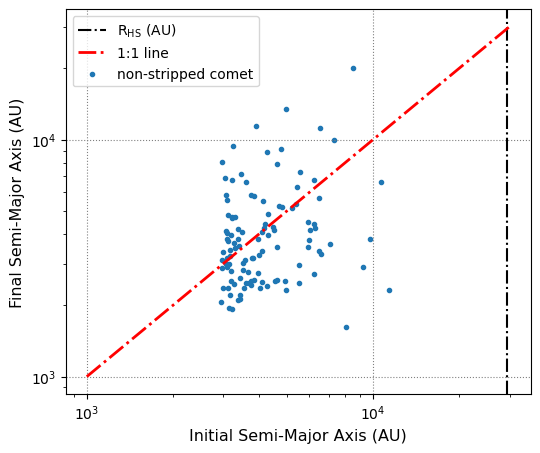}
  \includegraphics[width=\linewidth]{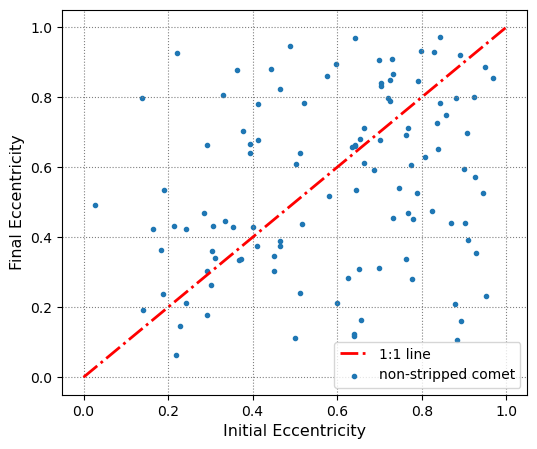}
  \caption{The initial versus final semi-major axis (top) and orbital eccentricity (bottom) of the remaining bound comets in one Oort cloud simulation around a host star with an escape time of 544 Myr and a maximum mass time of 50 Myr. These plots show how perturbations from other stars in the cluster have changed the orbits of the comets that remain in the Oort cloud when the star leaves the cluster. A 1:1 dashed line is plotted in red which allows us to observe which initial values are greater, equal to, and less than their final values. On the top plot, a vertical black dashed line marks the star's initial Hill sphere radius (R$_{\rm HS}$) of 29,400 AU, showing that the remaining comets have initial semi-major axes much smaller than this value. Comets that are stripped from the Oort cloud aren't shown in either panel.}
  \label{fig:fig3}
\end{figure}

Perturbations from passing stars can affect a comet’s orbital eccentricity and semi-major axis, as shown by the results in Figure \ref{fig:fig3}. Even if an Oort cloud keeps a high fraction of its comets when it leaves the cluster, it is possible for these remaining comets to have had their semi-major axis and/or eccentricity changed due to perturbations. Figure \ref{fig:fig3} shows the initial versus final semi-major axis and eccentricity of remaining bound comets in one simulation of a host star with an escape time of 544 Myr and an Oort cloud maximum mass time of 50 Myr (the simulation shown as a gray line in Figure \ref{fig:fig2}). The plot shows that all remaining bound comets have an initial semi-major axis smaller than the star's initial Hill sphere radius (29,400 AU). The comets were initially uniformly distributed with the Oort cloud's outer boundary approximately equal to the star's initial Hill sphere radius, meaning that all the comets with an initial semi-major axis close to the initial Hill sphere radius have been stripped from orbit due to perturbations. Some comets experienced little to no change in their semi-major axis while others experienced large changes, with these changes depending on the perturbations they each experience. To be specific, there are approximately 38.3\%  more bound comets with a final semi-major axis smaller than its initial value. Figure \ref{fig:fig3} additionally shows that every remaining bound comet ended with a different orbital eccentricity than it started with. This change indicates that each comet in the Oort cloud had its orbit perturbed at least slightly. This plot shows a wide variety of changes to the orbital eccentricities when the star leaves the cluster, with some comet orbits having their eccentricity increasing  (more elliptical) and some having their eccentricity decreasing (more circular). Specifically, there are approximately 11.3\% more bound comets that lost eccentricity than gained eccentricity. Overall, Figure \ref{fig:fig3} shows that almost all remaining bound comets have their final semi-major axis and orbital eccentricity differ from their initial values. This result indicates that even Oort clouds that leave the cluster with many comets remaining in orbit have different orbital characteristics than they did when the cloud reached its maximum mass. Although most comets experienced a change in their semi-major axis and eccentricity, the number of comets that experience an increase in either value is mostly compensated by close to an equal number that experience a decrease, thereby leaving the overall distributions mostly unchanged. Comparing the two distribution changes, we see that the eccentricity distribution remains closer to its original state than the semi-major axis distribution.

\subsection{Fraction of Comets Stripped}
\label{sec:results2}
\begin{figure}
  \includegraphics[width=\linewidth]{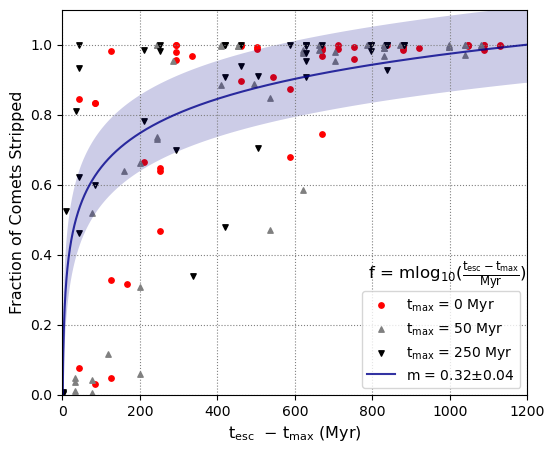}
  \includegraphics[width=\linewidth]{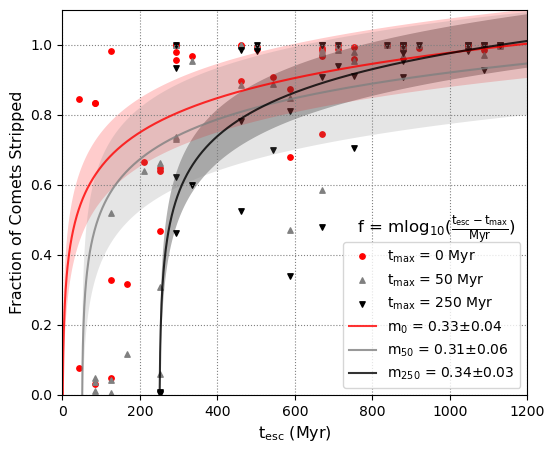}
  \caption{The fraction of comets stripped from each Oort cloud plotted against the number of years (t$_{ \rm esc}$ $-$ t$_{ \rm max}$) the Oort cloud spends in the cluster after it reaches its maximum mass (top) and the star's escape time  (t$_{\rm esc}$) (bottom). To represent the fraction of comets stripped over time, the top plot is fitted with one base ten logarithmic curve and the bottom plot is fitted with three base ten logarithmic curves, with the curves fitted to Oort clouds of each maximum mass time. The uncertainties are represented by shaded regions of error around the curves. Oort clouds that leave the cluster before their time of maximum mass are not considered in these plots.
  }
  \label{fig:fig4}
\end{figure}

 The example case discussed in \ref{sec:results1} is just the result of one out of the 50 host stars for the simulations. Each simulation experiences different changes made to the Oort cloud over time and one of the most important differences is the fraction of comets lost by the end. This result for the 50 sets of Oort cloud simulations is summarized in Figure \ref{fig:fig4}. This figure shows the fraction of comets stripped from orbit in each Oort cloud plotted against the time the Oort cloud spends in the cluster after its time of maximum mass (t$_{\rm esc}$ $-$ t$_{\rm max}$) and the time the star spends in the cluster (t$_{\rm esc}$).  

To represent the fraction of comets stripped (f) over time, the top plot is fitted with one base ten logarithmic curve fitted for all three maximum mass times following the equation:
\begin{equation}
        \rm f = m \log_{10} (\frac{t_{ \rm esc} - t_{ \rm max}}{Myr})
	\label{eq:quadratic}
\end{equation}
where m = 0.32 $\pm$ 0.04 and t$_{ \rm esc}$ $-$ t$_{ \rm max}$ is in Myr. This function has limits and is not valid for f $>$ 1 or f $<$ 0, as f = 1 means that all comets have been stripped from orbit and f = 0 means that no comets have been stripped.

The bottom plot is fitted with three base ten logarithmic curves that follow the same equation format as above, one for each maximum mass time with slopes of m$_0$ = 0.33 $\pm$ 0.04 for t$_{\rm max}$ = 0 Myr, m$_{50}$ = 0.31 $\pm$ 0.06 for t$_{\rm max}$ = 50 Myr, and m$_{250}$ = 0.34 $\pm$ 0.03 for t$_{\rm max}$ = 250 Myr. The uncertainties are represented by shaded regions of error around the curves. There is high variation as many points exist far outside the ranges of error, especially for early escape times and Oort clouds that spend less time in the cluster. Linear regression methods were used to determine these values of m by plotting log$_{\rm 10}$$\rm(\frac{t_{esc}-t_{max}}{Myr})$ against the fraction of comets stripped and fitting a straight line with a y-intercept at zero. It was decided that a single curve can best represent the trends in the top plot as the x-axis of t$_{\rm esc}$ $-$ t$_{\rm max}$ results in the three curves for each maximum mass time appearing almost identical within their uncertainty, meaning they can be represented well by just a single curve. In the bottom plot, the three curves for each maximum mass time appear separated due to the x-axis being solely t$_{\rm esc}$, so the shift is clear as Oort clouds that leave the cluster before their time of maximum mass are not represented in this plot. The values of m for the three curves are all within the uncertainty range of each other, indicating that the time an Oort cloud reaches maximum mass has little effect on the rate at which comets are stripped from the Oort cloud in the cluster after it has reached maximum mass. Although there is considerable variation, all curves show that the longer the Oort cloud remains in the cluster, the more comets are stripped from orbit with the fraction stripped increasing logarithmically with m = 0.32 $\pm$ 0.04 for each maximum mass time.  In terms of the specific variables we are considering, the fraction of comets stripped from orbit tends to increase along the curves as the host star's escape time and the time the Oort cloud spends in the cluster after it's maximum mass time increase. Furthermore, the curves show that regardless of the cluster's evolution, it usually takes roughly 1 Gyr for 90\% or more comets to be stripped from the host star, indicating that cluster evolution is a secondary factor when considering Oort cloud evolution in a clustered environment.

When comparing the individual Oort cloud data points in Figure \ref{fig:fig4}, the fraction of comets stripped from orbit is mostly higher for Oort clouds that reached maximum mass at 0 Myr (cluster's initial state) as more of them spend longer in the cluster and spend more time in higher density environments, meaning higher chances of close encounters with other stars. We consider an Oort cloud to have survived the cluster if most of its comets ($>$50\%) remain in orbit after its host star escapes. When the Oort clouds shown in Figure \ref{fig:fig4} as well as those not shown in Figure \ref{fig:fig4} (Oort clouds that leave the cluster before their maximum mass time) are all considered, we can compare percentages of Oort cloud survival. We found that if an Oort cloud in the star cluster reaches its maximum mass at the same time as the cluster's densest state (t$_{ \rm max}$ = 0 Myr), it tends to lose more than half of its comets as only 12\% of simulations where the Oort cloud has a maximum mass time of 0 Myr were able to keep more than half of their comets (all but one were around a star with t$_{ \rm esc}$ $<$ 200 Myr). Furthermore, 22\% of Oort clouds with a maximum mass time of 50 Myr were able to keep more than 50\% of their comets (all but one were around a star with t$_{ \rm esc}$ $<$ 250 Myr) and 32\% of Oort clouds with a maximum mass time of 250 Myr were able to keep more than 50\% of their comets (all but one were around a star with t$_{ \rm esc}$ $<$ 350 Myr). These survival percentages show that more Oort clouds with a maximum mass time of 250 Myr were able to keep more than half of their comets compared to the other maximum mass times. Additionally, these surviving Oort clouds are almost all those that spend relatively little time in the cluster with a t$_{ \rm esc}$ $<$ 200-350 Myr.

\begin{figure}
  \includegraphics[width=\linewidth]{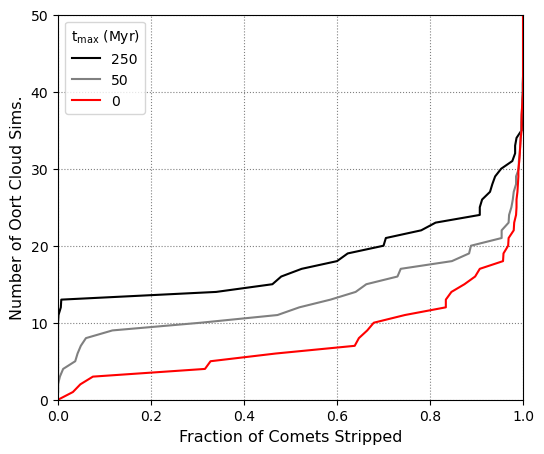}
  \caption{Cumulative distributions of the number of Oort cloud simulations that strip specific fractions of comets or less from orbit. These distributions show that more Oort clouds that reach maximum mass 250 Myr after their host star's formation can retain their comets.}
  \label{fig:fig5}
\end{figure}

\begin{table}
\centering
\begin{tabular}{||c c c c||} 
 \hline
 t$_{\rm max}$ 1 (Myr) & t$_{\rm max}$ 2 (Myr) & Statistic & P-Value \\ [0.5ex] 
 \hline\hline
 0 & 50 & 0.124 & 0.773 \\ 
 50 & 250 & 0.2 & 0.272 \\
 250 & 0 & 0.247 & 0.0748\\ [1ex] 
 \hline
\end{tabular}
\caption{Results of two-sided KS-tests of Figure 5's cumulative distributions. A p-value larger than the statistic value and greater than 0.05 indicates that the two groups being compared are likely from the same distribution. The table shows that only Oort clouds that reach maximum mass 0 and 250 Myr after the cluster's initial state are likely to belong to different distributions. }
\label{table:1}
\end{table}

To further explore comet loss due to perturbations, we consider the cumulative distributions of the number of Oort cloud simulations that strip a specific fraction of comets or less from orbit in Figure \ref{fig:fig5}. The results show that more Oort clouds lose more than half of their comets than Oort clouds that lose less than half. Additionally, the plot shows that when t$_{\rm max}$ is larger, a higher fraction of Oort clouds lose less than half of their comets. This result indicates that Oort clouds that are at their maximum mass at the star cluster's initial state are the most likely to lose all of their comets which we know from Figure  \ref{fig:fig4} is because most of these clouds spend more time in the cluster compared to the later maximum mass times. 

Two-sided Kolmogorov–Smirnov (KS) tests were done comparing the three cumulative distributions of the number of Oort clouds with different stripping fractions in Figure \ref{fig:fig5}. The test was done to see if the maximum mass time of an Oort cloud has any significant effect on its evolution in the star cluster. The KS-test results shown in Table 1 indicate that the survivability fraction distributions of only the 0 and 250 Myr maximum mass times could belong to separate distributions as their p-value is considerably smaller than their statistic value. Table 1 further shows that Oort clouds that reach maximum mass at 0 and 50 Myr belong to the most similar distribution as their p-value is much larger than 0.05 and the statistic value. This result is expected as the time gap between 0 and 50 Myr is much smaller than the time gap between 0 Myr and 250 Myr. Therefore, an Oort cloud that was at maximum mass at the star cluster's initial state is not significantly different in final mass than one that reaches its maximum mass 50 Myr later. However, an Oort cloud that reaches its maximum mass 250 Myr after the star cluster first forms tends to leave the cluster significantly more massive than those that form 250 Myr earlier as noticeably fewer comets are lost in the cluster.

\section{Discussion}
\label{sec:discussion}

The results of our simulations inform us about the potential evolution of Oort clouds around solar mass stars, including the Solar System's Oort cloud. Our model star cluster is comparable to a typical open cluster in a Milky Way-type galaxy. It has a similar density to the Sun's birth cluster \citep{Simon2009, Adams2010, Simon2019} with an initial density of 14 M$_\odot$/pc$^3$ and a comparable dissolution time. The escape time of the Sun from its birth cluster is not known, so comparing our Oort clouds around solar mass stars with a variety of escape times to predictions of the Sun's Oort cloud could give clues about where the Sun and its Oort cloud may have formed. Specifically, Figures \ref{fig:fig4} and \ref{fig:fig5} show that if the Sun had a late escape time from its birth cluster and an Oort cloud that reached maximum mass quickly after the Sun's formation,  then it is likely that most of its comets would be stripped from orbit before it leaves the cluster. \cite{Simon2021} argues that the majority of the Oort cloud formed after the Sun escaped its birth cluster as perturbations from nearby stars strip an earlier Oort cloud, resulting in only a small amount of comets from the original Oort cloud existing to this day. Our results tend to support this view as perturbations from passing stars cause most of our Oort clouds to lose more than half of their comets by the time they leave the cluster. However, some of our simulation results can support an alternate scenario of a quick Oort cloud build-up around a star with a very early escape time that can allow this primordial Oort cloud to survive to the present. 

It is estimated that there are 10$^{14}$ cometary objects currently in the Sun's Oort cloud. Rather than all these objects surviving from an initial formation in the Sun's birth cluster, our results indicate that usually, only part of an initial Oort cloud survives in the cluster, the surviving amount depending on when the Oort cloud reaches its maximum mass in the cluster and it's star's escape time. For a larger amount of the initial Oort cloud to survive the cluster, it must reach its maximum mass millions of years after the formation of the Sun's birth cluster or the Sun must have a relatively early escape time from the cluster (t$_{ \rm esc}$ $<$ 200-350 Myr) which can potentially allow it to keep most of a primordial Oort cloud until the present, assuming no other strong perturbations. 
 
\begin{figure}
  \includegraphics[width=\linewidth]{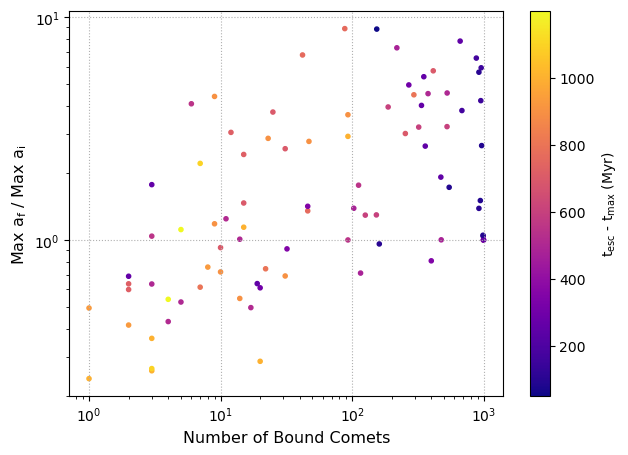}
  \caption{The ratio of the final maximum semi-major axis (a$_{\rm f}$) (AU) over the initial maximum semi-major axis (a$_{\rm i}$) (AU) of a comet in each Oort cloud plotted against the number of bound comets remaining in orbit when the Oort cloud leaves the cluster. The color bar signifies the amount of time the Oort cloud spends in the cluster ($\rm t_{\rm esc} - t_{\rm max}$) after its time of maximum mass. Oort clouds that lose all their comets when they leave the cluster have a final maximum semi-major axis of zero and are not shown in this plot. The figure has a slightly linear trend, as the ratio of max a$_{\rm f}$ over max a$_{ \rm i}$ increases, the number of bound comets tends to increase. Additionally, there is a trend in the colors that indicates that the greater max a$_{\rm f}$ is than max a$_{\rm i}$, and the greater the number of remaining comets, the less time the Oort cloud tends to have spent in the cluster after it reaches its maximum mass.}
  \label{fig:max_a}
\end{figure}

\begin{figure*}
\centering
  \includegraphics[width=\linewidth]{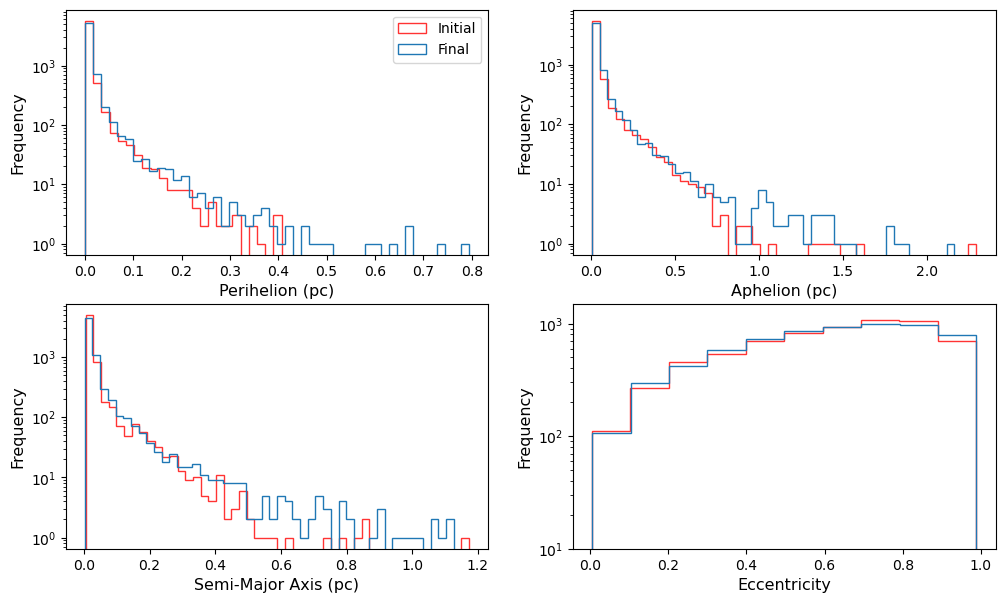}
  \caption{The initial and final perihelion (pc) (upper left), aphelion (pc) (upper right), semi-major axis (pc) (lower left), and orbital eccentricity (lower right) of all remaining bound comets in each Oort cloud simulation that loses at least half of its comets. Although perturbations from passing stars have resulted in the orbits of many comets in the Oort clouds to expand outwards, the final bound comets still orbit within the initial outer boundaries of all comets. The orbital eccentricity distribution of the remaining bound comets experienced little change, with the highest number of comets having an eccentricity of approximately 0.8.}
  \label{fig:e_a}
\end{figure*}

As explained in Section \ref{sec:maths2}, each Oort cloud has a different range of semi-major axes for its comet orbits as each host star has its own initial Hill sphere radius as the distance from the host to its closest star at formation is different for each star. The semi-major axis comparisons in Figure \ref{fig:max_a} and the semi-major axis, perihelion and aphelion distributions in Figure \ref{fig:e_a} show that some Oort clouds have comets with a maximum final semi-major axis (a$_{\rm f}$) larger than the maximum initial semi-major axis (a$_{\rm i}$) of all the comets, as well as a maximum final perihelion and aphelion larger than the maximum initial perihelion and aphelion. This result indicates that some Oort cloud comets moved to orbital distances beyond their star's initial Hill sphere radius. Other Oort clouds had the opposite scenario occur as their outer boundary shrank due to comet stripping or a decrease in the perihelion, aphelion, and/or semi-major axis of all remaining bound comets. We define remaining bound comets as comets that have remained in a bound orbit around their host star throughout the Oort cloud's time in the cluster and do not have an orbit that extends to distances beyond the star's Hill sphere radius at the time it leaves the cluster. It is important to note that a star's Hill sphere radius at the time it leaves the cluster can be as large as a few parsecs for certain stars that escape the least dense regions of the cluster due to their distant isolation from other stars. The average separation of stars in the Milky Way is estimated to be $\sim$1.5 pc, which is further than the aphelion of most of our simulated Oort cloud comets.

Figure \ref{fig:max_a} shows a slightly linear trend between the ratio of max a$_{\rm f}$ over max a$_{\rm i}$ and the number of bound comets remaining when the star leaves the cluster, with this ratio tending to increase as the number of comets increases. There is an additional trend that indicates that the greater the ratio between max a$_{\rm f}$ and max a$_{\rm i}$, and the greater the number of comets, the less time the Oort cloud tends to have spent in the cluster after it reaches its maximum mass. This result is understandable because spending less time in the cluster often results in fewer comets being stripped from orbit but more comets being perturbed to further distances while remaining in orbit. As additionally shown by the trend, spending more time in the cluster tends to result in a lower number of comets remaining in orbit and a larger max a$_{\rm i}$, as the comets that initially had large semi-major axes were stripped from orbit. There are also some outliers to these trends which can be explained by Oort clouds that spend less time in the cluster but experience more comet stripping and perturbations than average and Oort clouds that spend more time in the cluster but experience less stripping and perturbations than average. For example, an Oort cloud that forms around a star with an early escape time that reaches its maximum mass in a region of the cluster that is denser than average could experience stronger perturbations than expected.

 Simulations of stars with initial Hill sphere radii close to the upper boundary of the outer Oort cloud could represent the Sun's Oort cloud if we assume it had boundaries at its maximum mass similar to the present. It is also possible that the Oort cloud's boundaries were different at its maximum mass as shown in Figure \ref{fig:max_a}. Only two of our host stars had an initial Hill sphere radius greater than 100,000 AU, indicating that it could be rare for a solar mass star in this cluster to have a Hill sphere radius at formation as large as 200,000 AU. Figure \ref{fig:max_a} shows that some Oort cloud simulations have a comet with a larger maximum final semi-major axis (specifically some are close to 200,000 AU although this is not visible in the figure), showing that the outer boundary of these Oort clouds has spread to distances beyond it's star's initial Hill sphere radius by the time it leaves the cluster. Similarly to these simulations, the outer boundary of the Solar System's Oort cloud could have increased over time as the Sun remained bound to its birth cluster due to the perturbations caused by passing stars. The outer boundary of the Oort cloud could have also grown to its current distance due to the influence of the Galactic total field which is not explored in our simulations. Additionally, it could be possible that the Sun's Oort cloud continued to capture material from other stars' Oort clouds while still in the cluster after it reached its maximum mass, but this is also not considered in our simulations. \cite{Simon2021} found that half the material in the Solar System region between 100-10,000 AU and a quarter of the material in the $>$10,000 AU region could be captured from free-floating debris in the star's birth cluster or from the protoplanetary disk of other stars in the cluster.

The two host stars with initial Hill sphere radii closest to the present-day Oort cloud's outer boundary have an initial Hill sphere radius of approximately 281,702 AU and 120,036 AU. Both of these stars escaped the cluster very early with the first one escaping at 84 Myr and the second one escaping at 126 Myr. By the time the star with the larger Hill sphere radius escaped the cluster, 83.4\% of comets were stripped for the Oort cloud with t$_{\rm max}$ =  0 Myr, and 4.9\% of comets were stripped for the one with t$_{\rm max}$ = 50 Myr. By the time the star with the smaller Hill sphere radius escaped the cluster, 98.1\% of comets were stripped for the Oort cloud with t$_{\rm max}$ = 0 Myr, and 51.9\% of comets were stripped for the one with t$_{\rm max}$ = 50 Myr. The percentages show that if an Oort cloud with an initially large Hill sphere radius reached maximum mass immediately after its star's formation, even with an early escape time, it still loses the majority of its comets, with the second star's Oort cloud losing more than 90\% of its mass. The only Oort cloud that was able to retain most of its comets was the one that took 50 Myr to build up to maximum mass around a star with an escape time of 84 Myr. This result indicates that if an Oort cloud built up to its maximum mass around a star with a large initial Hill sphere radius while it was still in its birth cluster and survived to the present day, then it likely reached this maximum mass very close to its star's escape time, allowing it to spend little time in the cluster and lose only a few comets. Furthermore, these stars left the cluster with Oort clouds that do not closely match the estimations of the Sun's Oort cloud's boundaries, even though they initially had somewhat close to these values. These findings support the idea that the Oort cloud does not begin with its present-day size or boundaries in its star's birth cluster as it would experience many noticeable changes over time due to perturbations from passing stars (and other factors that weren't considered in our simulations).

To explore the change in orbital properties of remaining bound comets, we show in Figure \ref{fig:e_a} histograms of their initial and final perihelion, aphelion, semi-major axis, and eccentricity for Oort clouds that lose at least 50\% of their initial comets.  Figure \ref{fig:e_a} shows that perturbations from passing stars have resulted in some comet orbits expanding outwards as their final aphelion, semi-major axis, and/or perihelion increased to larger distances. Although the orbits of some comets in the Oort clouds expand outwards, Figure \ref{fig:e_a} also shows that the final remaining comets still orbit within the initial boundaries of all Oort clouds. We investigate the changes in aphelion and perihelion as an increased semi-major axis does not necessarily mean that a comet orbit expanded further from its star if its aphelion moved closer to it. 

The initial semi-major axis range of all comets from all Oort clouds combined is approximately 1,000-280,000 AU with a minimum perihelion of 95 AU and maximum aphelion of 540,000 AU. The Oort cloud semi-major axis, perihelion, and aphelion ranges shown in Figure \ref{fig:e_a} are larger than what is predicted for the Sun's Oort cloud, but this plot shows the remaining bound comets from all Oort cloud simulations that lose more than 50\% of their comets, so all comets in the Sun's Oort cloud could easily fit on this plot within these boundaries. 

The maximum initial aphelion of the remaining bound comets of nearly 540,000 AU (with a $\sim$280,000 AU) can be thought of as the initial outermost point of all the Oort clouds shown in Figure \ref{fig:e_a}, as none orbit a star further than this point. The maximum final aphelion of the remaining bound comets is approximately 450,000 AU, which is slightly smaller than the initial maximum aphelion of all comets. This decrease in maximum aphelion indicates that perturbations caused the initially most distant comet orbit to shift closer to its star without being stripped from the system, likely perturbing its orbit before the comet ever reached its initially distant aphelion. However, as seen in Figure \ref{fig:e_a}, quite a few comet orbits do extend further than they did initially, causing the distribution of comets to increase at further distances without extending past the initial outer boundary of all Oort clouds. This result provides further evidence that perturbations can cause some Oort clouds to expand to larger distances as their star enters less clustered environments. Although Figure \ref{fig:e_a} shows some Oort cloud comet orbits expanding to further distances, it also shows that the overall distribution of the final bound comets is still similar to the initial distribution. 

A few comet orbits crossed the innermost point of all Oort clouds as their perihelion greatly decreased. Specifically, the minimum initial perihelion of the remaining bound comets is approximately 95 AU (with a = 2,381 AU), which can be thought of as the initial innermost point of all the Oort clouds, as none orbit the star closer than this point. The minimum final perihelion of the remaining comets is approximately 16 AU (with a = 1,220 AU). This perihelion value is noticeably smaller than the minimum initial perihelion of the comets, indicating that it orbits at distances much closer to the star. Perturbations from passing stars have caused the orbit of this comet (and a few others) to leave the Oort cloud region entirely, crossing the inner boundary of the Oort cloud at maximum mass and reaching planet orbit regions. Although no planets were included in our simulations, it is important to note that if any planets were included, the orbital trajectory of this comet could be heavily influenced by their gravity. 

In the initial and final state of the Oort cloud simulations, it was observed that the comets with the largest semi-major axes tend to have more elliptical orbits while the rest of the comets have orbits of a wider range of eccentricities. Figure \ref{fig:e_a} shows that the overall orbital eccentricity distribution of the remaining bound comets practically did not change. The eccentricity of the comet orbits is various, but there is a larger number of comets that have more eccentric orbits, both initially and finally, with the highest number of comets having an eccentricity of approximately 0.8. Therefore, perturbations from passing stars in a cluster do not change the overall eccentricity distribution of remaining bound comets but it does have some effect on their semi-major axis, perihelion, and aphelion distributions.  
\\
\\
\section{Conclusions}
\label{sec:conclusions}
 We set out to explore how the amount of time a star spends in its birth cluster and the time its Oort cloud reaches its maximum mass affect the Oort cloud's ability to survive in the cluster using $N$-body simulations. After evolving a star cluster with an initial density of 14 M$_\odot$/pc$^3$ to dissolution, 50 stars of 1 M$_\odot$ with a range of cluster escape times were chosen to host Oort clouds composed of 1000 comets at their time of maximum mass. For each host star, we considered Oort clouds that reach a maximum mass 0, 50, and 250 Myr after the star cluster's formation, and simulated their evolution in the cluster starting from these three times. The main results of the 150 Oort cloud simulations showed that for a star cluster with an initial density of 14 M$_\odot$/pc$^3$ : 
 
 \begin{itemize}
  \item  When the fraction of comets that are stripped from each Oort cloud is plotted against t$_{ \rm esc}$ $-$ t$_{ \rm max}$ in Myr, the points can be fit with a single base ten logarithmic curve with a slope constant of m = 0.32 $\pm$ 0.04. This relationship indicates that the longer the Oort cloud remains in the cluster, the more comets are stripped from orbit with the fraction stripped increasing logarithmically at approximately the same rate for each t$_{ \rm max}$. 
  \item 12\% of Oort Clouds that are at maximum mass at the time of the star cluster's densest/initial state keep at least 50\% of their comets when they leave the cluster. This small percentage of surviving Oort clouds is almost all those with host stars that escaped the cluster early ($<$200 Myr). On average, the longer an Oort cloud spends in the cluster after it has reached maximum mass, the more comets are stripped from orbit. Thus, an Oort cloud that is at maximum mass at the star cluster's initial state but escapes the cluster a few years later could retain most of its comets. 
  \item 22\% of Oort clouds that build up to a maximum mass 50 Myr after the initial state of the star cluster retain more than half of their comets. All but one of these surviving Oort clouds formed around a star with a cluster escape time less than 250 Myr.
  \item 32\% of Oort clouds that build up to a maximum mass 250 Myr after the initial state of the star cluster retain more than half of their comets. All but one of these surviving Oort clouds formed around a star with a cluster escape time less than 350 Myr. 
  \item When all simulations are considered, it doesn't significantly matter if an Oort cloud reaches maximum mass 0 or 50 Myr after the star cluster's formation because it does not change how the Oort cloud evolves in the cluster, but an Oort cloud that reaches maximum mass 250 Myr later evolves differently than an Oort cloud that is at maximum mass at the star cluster's initial state. 
\end{itemize}

We conclude that if the Sun was born in a clustered environment, as suggested in \citealp{Simon2009}, and many other studies (e.g. \citealp{Gaidos1995, Adams2001, Looney2006, Adams2010}), then a surviving Oort cloud likely reached maximum mass greater than 50-250 Myr after the formation of the Sun's birth cluster and escaped the cluster less than 100-200 Myr later. Our simulation results overall showed that in a star cluster with an initial density of 14 M$_\odot$/pc$^3$, Oort clouds that reach maximum mass at the star cluster's initial state lose on average 85.1\% of their comets. In contrast, Oort clouds that reach maximum mass 50 Myr and 250 Myr after the initial cluster state lose on average 74.6\% and 64.9\% of their comets, respectively. 
    
To summarize, analyzing our simulations in the context of the Sun's Oort cloud indicate:
\begin{itemize}
  \item It could be rare for a solar mass star to have a large Hill sphere radius at formation, as it is rare for a star to be born in the outskirts of a cluster and remain there until the cluster dissolves, indicating that if the Oort cloud is at maximum mass at the formation of the Sun's birth cluster, then the Sun may have formed in a less dense cluster or the Oort cloud was initially much more compact than it is predicted to be today. 
  \item If the present-day Oort cloud did form around the Sun while it was still in its birth cluster, most of the Oort cloud's materials would have been initially located in much more compact orbits, but during its co-evolution with the Sun's birth cluster, perturbations from passing stars injected energy and angular momentum, which drove the expansion of the Oort cloud. 
\end{itemize}

To expand on our research on Oort clouds, the next steps would be to simulate Oort cloud evolution around various types of stars within a diverse range of star clusters. Exploring these simulations would allow the relationship between star type and Oort cloud, and star cluster and Oort cloud to be developed. 
\\

The authors would like to thank the University of Toronto’s 2022 Summer Undergraduate Research Program and acknowledge financial support from the Natural Sciences and Engineering Research Council of Canada and an Ontario Early Researcher Award, which helped make this research possible.

The simulation data presented in this paper can be made available upon request.
\\
\\
\\
\bibliography{refs}{}
\bibliographystyle{aasjournal}



\end{document}